\documentclass{amsart}
\usepackage{amssymb,amsmath}
\usepackage{amscd,cite}
\newtheorem{theorem}{Theorem}

\newtheorem{proposition}{Proposition}

\newtheorem{corollary}{Corollary}

\newcommand{\Rset}{\mathbb{R}}

\newcommand{\hcA}{\hat{\mathcal{A}}}
\newcommand{\cA}{\mathcal{A}}

\newcommand{\cG}{\mathcal{G}}
\newcommand{\cT}{\mathcal{T}}
\newcommand{\cJ}{\mathcal{J}}
\newcommand{\cQ}{\mathcal{Q}}

\newcommand{\cP}{\mathcal{P}}

\newcommand{\miI}{I}
\newcommand{\miJ}{J}

\newcommand{\linopspace}{{\mathcal{L}}}
\newcommand{\opalgebra}{{\mathcal{T}}}

\newcommand{\supth}{^{\mathrm{th}}}

\newcommand{\pr}{\operatorname{pr}}

\newcommand{\wt}{\operatorname{wt}}
\newcommand{\spn}{\mbox{span}}

\newcommand{\tu}{{\tilde{u}}}

\newcommand{\rkmin}{r_{\text{min}}^{(k)}}
\newcommand{\rkmax}{r_{\text{max}}^{(k)}}

\begin{document}
\title[differential operators admitting invariant polynomial
subspaces]{Structure theorems for linear and non-linear
differential operators admitting invariant polynomial subspaces}
\author{David Gomez-Ullate}
\address{ Dep. Matem\`atica Aplicada I, Universitat Polit\`ecnica de Catalunya, ETSEIB, Av. Diagonal 647, 08028 Barcelona, Spain. }
\author{ Niky Kamran }
\address{Department of Mathematics and Statistics, McGill University
Montreal, QC, H3A 2K6, Canada}
\author{Robert Milson}
\address{Department of Mathematics and Statistics, Dalhousie University, Halifax, NS, B3H 3J5, Canada}
\begin{abstract}
 In this paper we derive structure
theorems that characterize the spaces of linear and non-linear
differential operators that preserve finite dimensional subspaces
generated by polynomials in one or several variables. By means of
the useful concept of deficiency, we can write explicit basis for
these spaces of differential operators. In the case of linear
operators, these results apply to the theory of quasi-exact
solvability in quantum mechanics, specially in the multivariate
case where the Lie algebraic approach is harder to apply. In the
case of non-linear operators, the structure theorems in this paper
can be applied to the method of finding special solutions of
non-linear evolution equations by nonlinear separation of
variables.
\end{abstract}
\maketitle

\section{Introduction}
It is a fact that the Schr\"odinger operators whose point
spectrum, or at least part of it, can be computed algebraically
are often related to differential operators admitting invariant
spaces of polynomials. Lie algebras have played a unifying role in
this area, because many of these polynomial spaces turn out to be
irreducible modules for a faithful representation of a
finite-dimensional Lie algebra by means of first-order
differential operators. The classical theory of quasi-exactly
solvable potentials has thus been built on the assumption that the
exactly solvable Schr\"odinger operator under study should be
expressible as a quadratic element in the universal enveloping
algebra of a finite-dimensional Lie algebra of first-order
differential operators, admitting an explicitly computable
invariant subspace of square-integrable functions, or a complete
infinite flag thereof \cite{ST,T,U1,KO}.
 Burnside's Theorem serves as a strong
argument in favor the Lie algebraic approach since it implies that
any endomorphism of an irreducible module for a Lie algebra can be
represented as a polynomial in the generators of the algebra.
However, recent results show that the Lie algebraic approach
suffers from various limitations that reduce severely its
applicability:
\begin{enumerate}
\item In the case of polynomial subspaces in one variable, the Lie
algebraic approach can only be applied to find the differential
operators that leave the polynomial space $\mathcal P_n=\langle
1,z,z^2,\dots,z^n \rangle$ invariant, but it cannot characterize
the set of differential operators that map $\mathcal P_n$ into
$\mathcal P_m \subset \mathcal P_n$ with $m<n$. This simple
problem has motivated the important notion of {\em deficiency}
used throughout this paper, and applications of it can be found in
the construction of solvable classical many-body problems by
considering the motion of zeros of polynomials whose coefficients
evolve in a controlled manner \cite{Cal_book}. \item Other
subspaces generated by polynomials exist which are not isomorphic
to $\mathcal P_n$. The Lie algebraic approach cannot be applied in
these cases, as was already noted by Post and Turbiner, who
characterized the spaces of linear differential operators which
preserve polynomial subspaces in one variable generated by
monomials. In their work \cite{PT} they solved this problem with
no reference to Lie algebras. The case of a general space spanned
by polynomials
--- referred to as the {\em generalized Bochner problem} in
\cite{PT}--- remains still open. Somewhat surprisingly this direct
approach has not been pursued until very recently, where it has
been shown that the class of quasi-exactly solvable potentials is
larger than the Lie-algebraic class \cite{GKM3}, and that even in
Lie-algebraic potentials other non-$\mathfrak{sl}_2$
algebraizations exist which allow to obtain more levels from the
energy spectrum of the Hamiltonian,\cite{GKM4}. The existence of
differential operators that preserve a general polynomial space
and cannot be expressed as quadratic combinations of the
generators of $\mathfrak{sl}_2$ is not in contradiction with
Burnside's theorem, since a general polynomial space is not the
carrier space for an irreducible representation of
$\mathfrak{sl}_2$.

\item In the case of multi-variable polynomial subspaces, the
problem of characterizing the set of linear differential operators
that leave these spaces invariant becomes untractable in the Lie
algebraic approach. The reason is that, contrary to the single
variable case where essentially $\mathfrak{sl}_2$ is the only
algebra of first order differential operators with finite
dimensional representations, in more variables many more algebras
exist. But moreover, the characterization of second order
operators as quadratic combinations of the generators of these
algebras requires an extensive analysis of the syzygies
corresponding to the primitive ideals associated to the
irreducible representations~\cite{FK}. These problems are entirely
bypassed in the direct approach, as shown in Section 3 of this
paper, where a simple characterization is given  for the set of
linear differential operators of any given order $r$ that leave
the simplicial module
\[
{\mathcal{P}}_{n}=\mbox{span}\{x_{1}^{i_{1}}\ldots
x_{N}^{i_{N}}\,|\,i_{1}+\cdots +i_{N}\leq n\}
\]
invariant. Our results coincide with the formulas  for the special
case $N=2$ and $r=2$ derived in \cite{FK}  using the Lie algebraic
method.
\end{enumerate}
It has now  become clear that the connection to Lie algebras is
not an essential feature of exact or partial solvability. Our goal
in the first two Sections of this paper is to present a direct
method to characterize linear differential operators with
invariant polynomial subspaces which is simpler and more powerful
than the Lie algebraic approach. We restrict in this paper to the
simplest case of polynomial subspaces, namely the simplicial
modules $\mathcal P_n$, the case of general polynomial subspaces
shall be treated elsewhere. For any number of variables $N$, we
provide an explicit basis for the space of linear differential
operators of any order $r$ that map $\mathcal P_n$ into $\mathcal
P_m \subseteq \mathcal P_n$ with $m\leq n$. It should be stressed
that although these results allow to construct many differential
operators with invariant finite dimensional polynomial subspaces,
in general it is not known whether a transformations exists that
puts the operator in Schr\"odinger form. Therefore the results in
Sections 2 and 3 are only a first necessary step in the theory of
higher dimensional quasi-exact solvability. A general theory would
need to face the difficulties of the equivalence problem \cite{M}.
Despite this fact, it is worth mentioning that a few examples of
partially solvable multi-dimensional Hamiltonians exist
\cite{tur98,GGR00,FGGRZ01}, which are mostly extensions of the
Calogero-Sutherland class.

This paper also addresses the study of nonlinear differential
operators with polynomial nonlinearities which possess invariant
polynomial subspaces. The motivation for this study is twofold:
\begin{enumerate}
\item In \cite{KMO}, the important concept of {\em operator duals}
is introduced. Given a finite dimensional space of functions
$\mathcal F_n= \mbox{span}\{f_1,\dots,f_n\}$ which are required to
satisfy certain regularity conditions, the operator duals are
linear differential operators defined by the relations
\[D_i[f_j]=\delta^i_j.\]
These operators are used in the reduction of non-linear evolution
equations to dynamical systems by a method of non-linear
separation of variables. In \cite{KMO} the existence of these
differential operators is proved together with results on the
regularity of the coefficients. The proof is constructive and
therefore given any space $\mathcal F_n$ whose basis elements
satisfy the required regularity conditions, non-linear evolution
equations can be written which have solutions in the space
$\mathcal F_n$, i.e. special solutions exist of the form
$u(t,x)=\sum_{i=1}^n c_i(t) f_i(x)$ where the coefficients
$c_i(t)$ satisfy a system of coupled non-linear ODEs. However, the
order of the operator duals is precisely the dimension of the
space $\mathcal F_n$. The price to pay in order to have invariant
spaces of high dimension is that the order of the resulting
evolution equations grows with the dimension of the space. For
dimensions $n$ larger than six this reduces the applicative
interest of the resulting equations. The motivation coming from
\cite{KMO} is to construct operators that generalize the operator
duals, so that the order of the resulting equation and the
dimension of the invariant subspace can be independently chosen.
In the case of polynomial subspaces $\mathcal P_n$ , the
generalization of the operator duals to arbitrary order $r\leq n$
are the so called {\em maximal deficiency operators} introduced in
Section 2.

\item The second motivation comes also from the results on
non-linear separation of variables by King, Galaktionov and
Svirshchevskii \cite{Sv,Ga,K1,K2,K3}. In these papers special
interest is given to translation-invariant evolution equations
with quadratic non-linearities which admit solutions via
non-linear separation of variables. From a physical context,
applications are found in nonlinear diffusion and thin film
equations. In this paper we extend these results by providing a
comprehensive structure theory for autonomous nonlinear  operators
that preserve a polynomial space $\mathcal P_n$.
\end{enumerate}

Our paper is organized as follows. In Section 2, we present our
direct approach in the case of linear differential operators in
one variable. Besides the order of a differential operator, we
introduce two key invariants which can be freely specified, which
are the degree and the deficiency of the operator relative to a
polynomial space $\mathcal P_n$. The order, degree and deficiency
are shown to specify the operator uniquely up to scaling by a
constant. The operators of maximal deficiency generalize the
operator duals of ~\cite{KMO} to any order lower than $n$. We give
an explicit basis for the space of operators of given order and
deficiency. Section 3 is concerned with linear differential
operators in $N$ variables, where all the results of Section 2 are
shown to extend to the case of simplicial modules, that is
multivariate polynomials of total degree bounded by a given
integer. Section 4 is concerned with non-linear operators
preserving polynomial modules, where we give an explicit
decomposition theorem for the most general non-linear operator
with polynomial coefficients preserving a simplicial module.
Section 5 studies the deficiency concept for non-linear operators
with polynomial nonlinearities. Section 6 concentrates on
non-linear operators that are translation invariant (autonomous)
deriving also structure theorems for this class. On Section 7 the
application of these results to non-linear separation of variables
is discussed while some explicit formulas are given for
quadratically non-linear autonomous operators in Appendix A.

\section{Linear operators in one variable}

In this section, we consider the class of scalar linear
differential operators on the real line, with polynomial
coefficients. We are interested in the subclass of operators which
have a definite {\em order}, {\em degree} and {\em deficiency}.
These quantities are defined as follows.
The order of
\begin{equation}
\label{eq:linode} L=\sum_{i=0}^{r}a_{i}(x)D^{i},\quad
D:=\frac{d}{dx},
\end{equation}
is as usual the largest $r$ for which the coefficient $a_{r}(x)$
is not identically zero. We say that $L$ is of degree $d\in
\mathbb{Z}$ if for all $j\in \mathbb{N}$, there exists $c_{j}\neq
0 \in \mathbb{R}$ not all zero such that
\begin{equation}
L[x^{j}]=c_{j}\,x^{j+d}.
\end{equation}
In order to define the deficiency, we fix $n\in \mathbb{N}$ and
consider the vector space
\begin{equation}
{\mathcal{P}}_{n}=\spn\{1,x,\ldots,x^{n}\}
\end{equation}
of polynomials in $x$ of degree less than or equal to $n$. We say
that $L$ has {\em deficiency $m \in \mathbb{Z}$ relative to
$\cP_n$} if
\begin{equation}
L{\mathcal{P}}_{n}\subset {\mathcal{P}}_{n-m},\;\mbox{but }
L{\mathcal{P}}_{n}\not\subset {\mathcal{P}}_{n-m-1}.
\end{equation}
Let ${\linopspace}_{r,m}^{(n)}$ denote the set of linear differential
operators with polynomial coefficients, of order less than or equal to
$r$ and of deficiency greater than or equal to $m$ relative to
$\mathcal{P}_n$.  Again, we emphasize that
the notion of  operator deficiency only makes sense relative to a particular
$n$.  Most of our discussion will be carried out with the assumption
that the $n$ in $\mathcal{P}_n$ has been fixed.  As such, we will often
omit the $n$ in our terminology and notation,  simply speak of the
deficiency of an operator, and write $\linopspace_{r,m}$
instead of  ${\linopspace}_{r,m}^{(n)}$.

\begin{proposition}
  The set ${\linopspace}_{r,m}$ is a subspace of the vector space of
  all linear differential operators, i.e., it is closed under linear
  combinations.
\end{proposition}
\noindent\noindent\textbf{Proof}
  Given linear operators $L, L'\in\linopspace_{r,m}$, the order of
  any
  linear combination of $L$ and $L'$ is less than or equal to $r$.
  Similarly, the deficiency of a linear combination is greater than or
  equal to $m$. \qed

\begin{proposition}
  \label{prop:deford}
  The deficiency of a non-zero linear operator cannot exceed its
  order.
\end{proposition}
\noindent\textbf{Proof}
  Suppose that $L\in\linopspace_{r,m}$ is a linear operator such that
  $m>r$.  The operator $D^{n-m+1}L$ annihilates $\mathcal{P}_{n}$, but has
  order less than $n+1$.  This is impossible. \qed

Trivial examples of operators with given order, degree and
deficiency are given by the operator $D^{i}$, which has order $i$,
degree $-i$ and deficiency $i$, and the multiplication operator
$x^{j}$ has order zero, degree $j$ and deficiency $-j$. These
operators do not depend on the degree $n$ of the polynomial space
${\mathcal{P}}_{n}$. A more significant example, which depends
explicitly on $n,r,m,d$ with $0\leq m\leq r\leq n$ and $-m\leq
d\leq r-m$ is the operator
\begin{equation}
\label{eq:Tdef} L_{rmd}:=x^i\, (\,n-j-xD\,)_{k}D^{j},\quad
i=r-m,\; j=r-d-m,\; k=d+m
\end{equation}
where we have introduced the Pochhammer operator
\begin{equation}
\label{eq:poch}
(\, n-xD \,)_{k}:=(-1)^{k}(\, xD-n \,)(\, xD-(n-1)\,)\cdots(\,
xD-(n-k+1)\,).
\end{equation}
A  basic result is
the following:
\begin{proposition}
  \label{prop:Trmdinfo}
  The operator
  $L_{rmd}$ has order $r$, degree $d$ and deficiency $m$.
\end{proposition}
\noindent\textbf{Proof}
  Since the $k$-fold composition appearing in the right-hand-side of
  ~(\ref{eq:poch}) annihilates the monomials $x^{n},\ldots,x^{n-k+1}$,
  the operator in \eqref{eq:poch} has order $k$, degree zero and
  deficiency $k$.  It follows that the operator
  $(\,n-j-xD\,)_{k}D^{j}$ has order $r=k+j$,  deficiency $m=k+j$ and degree
  $d=-j$.  By left-multiplying by a monomial of $x$ we raise the
  degree and lower the deficiency, so that $L_{rmd}$ has order
  $r$, deficiency $m$, and degree $d$. \qed

Following Proposition \ref{prop:deford}, it is helpful to refer to an
operator whose deficiency equals its order, as a \emph{maximal
  deficiency operator}.  To this end, we will use the symbol
\begin{equation}
  K_{rj} =   \frac{1}{(r-j)!} L_{r,r,-j}=
  \frac{1}{(r-j)!} (\,n-j-xD\,)_{r-j}D^{j},\quad 0\leq j\leq r\leq n,
\end{equation}
to denote the operators of maximal deficiency.  The normalization
constant of $\frac{1}{(r-j)!}$ will be useful in later formulas.
These operators have a number of interesting properties, and play a
key role in our theory.
\begin{proposition}\label{prop:uniquelinear}
  Up to multiplication by a non-zero real constant, the operator
  $K_{rj}$ is the unique $r\supth$ order maximal deficiency operator
  with polynomial coefficients having degree $d=-j$, where $0\leq
  j\leq r$.
\end{proposition}
\noindent\textbf{Proof}
  Let $L$ be an $r\supth$ order maximal deficiency operator with
  polynomial coefficients and having degree $d$.  Since $L$ has
  polynomial coefficients, $d\geq -r$.  As well, $L$ maps $x^k$ to a
  multiple of $x^{k+d}$.  Hence,
  \begin{equation*}
    L[x^k] = 0,\quad  n-r-d< k \leq n.
  \end{equation*}
  However, a non-zero $r\supth$-order operator can annihilate at most
  $r$ monomials.  Thus, we have established that $0\leq j\leq r$,
  where $j=-d$.

  The leading order term of both $L$ and $K_{rj}$ is a multiple of
  $x^{r-j} D^r$.  Hence, there exists an $a\in\mathbb{R}$ such that
  the order of $L-a K_{rj}$ is less than $r$. However, the deficiency
  of $L-a K_{rj}$ is greater than, or equal to $r$.  Therefore, by
  Proposition \ref{prop:deford}, we have $L=aK_{rj}$. \qed

\begin{proposition}
  \label{prop:Trjproduct}
  For every $0\leq j\leq r\leq n$, we have
  \begin{equation}\label{eq:Trjproduct}
    (r-j)!\,K_{rj}= (\,n-j-xD\,)_{r-j}\, D^{j} = D^j (\,n-xD\,)_{r-j}.
  \end{equation}
\end{proposition}
\noindent\textbf{Proof}
  The operators $(\,n-j-xD\,)_{r-j}\, D^{j}$ and $D^j
  (\,n-xD\,)_{r-j}$ both have order $r$, deficiency $r$ and degree
  $-j$.  By Proposition \ref{prop:uniquelinear} they differ by a
  scalar multiple.  By comparing the coefficients of the leading
  order, we see that these two operators are actually equal. \qed
\begin{proposition}
  \label{prop:Trjrecdef}
  For a fixed $r$, the operators $K_{rj}$ are recursively defined
  by
  \begin{equation}
    \label{eq:Trjrecdef}
    [D,K_{rj}] = - K_{r,j+1},\qquad K_{rr} = D^r.
  \end{equation}
\end{proposition}
\noindent\textbf{Proof}
    Setting $k=r-j$, we have, by Proposition \ref{prop:Trjproduct},
  \begin{align*}
    [D, K_{rj}] &= \frac{1}{(r-j)!}\, \left(D^{j+1} (n-xD)_k -
      (n-j-xD)_k D^{j+1} \right)\\
    &=   \frac{1}{(r-j)!} \,D^{j+1}((n-xD)_k- (n+1-xD)_k) \\
    &= -\frac{k}{(r-j)!}\, D^{j+1} (n-xD)_{k-1}\\
    &=- K_{r,j+1}.\quad \square
  \end{align*}
\noindent
We can also expand the maximal deficiency operators into an operator
sum.
\begin{proposition}
  \label{prop:Krjsum}
  We have
  \begin{equation}
    \label{eq:Krjsum}
    K_{rj}=   \sum_{k=0}^{r-j} (-1)^{k} \binom{n-k-j}{n-r}
    \frac{x^k}{k!}\, D^{k+j}
  \end{equation}
\end{proposition}
\noindent\textbf{Proof}
  Let $\hat{K}_{rj}$ denote the right hand side of \eqref{eq:Krjsum}.
  A direct calculation shows that
  \begin{equation*}
    [D , \hat{K}_{rj}] = - \hat{K}_{r,j+1},\qquad \hat{K}_{rr} =D^r.
  \end{equation*}
  Hence, by Proposition \ref{prop:Trjrecdef}, $K_{rj}$ and
  $\hat{K}_{rj}$ have the same recursive definition.  Therefore,
  the two operators are equal. \qed

\begin{proposition}
  \label{prop:maxdef}
  We have $\dim \linopspace_{r,r}=r+1$.  Indeed, the operators $K_{rj}$,
  $j=0,1,\dots,r$  form a basis for the vector space $\linopspace_{r,r}$.
\end{proposition}
\noindent\textbf{Proof}
  Let $L\in \linopspace_{r,r}$ be given.  Decomposing $L$ into operator
  monomials of homogeneous degree, we have by Proposition
  \ref{prop:deford} and the fact that $\cP_n$ is generated by
  monomials,
  \begin{equation*}
    L=\sum_{j=0}^r a_j\, K_{rj}.
  \end{equation*}
  Since the $K_{rj}$ are linearly independent, we conclude that $\dim
  \linopspace_{r,r}=r+1$. \qed

\noindent The following corollary follows immediately.
\begin{proposition}
  \label{prop:Trm}
  We have
  \begin{equation}
    \label{eq:defrel}
    {\linopspace}_{r,m}={\mathcal{P}}_{r-m}\otimes
    {\linopspace}_{r,r},\quad\mbox{with}\quad
    \dim \linopspace_{r,m} = (r+1)(r-m+1).
  \end{equation}
  In particular, the set of operators of order $r$ or less that map
  ${\mathcal{P}}_{n}$ to itself is of dimension given by
  $(r+1)^{2}=\dim \operatorname{gl}(r+1,\mathbb{R})$.
\end{proposition}
\noindent We conclude this section with a simple example:

\noindent\noindent\textbf{Example 1.}\label{ex1} A basis for the
vector space ${\linopspace}_{2,2}$ of operators of order two and
deficiency two is given by
\begin{eqnarray*}
K_{22}&=& D^{2},\\
K_{21} &=& (\,xD-n+1\,)D =xD^{2}+(1-n)D,\\
K_{20}&=&(\,xD-n\,)(\,xD-n+1\,)=x^{2}D^{2}+2(1-n)\,xD+n(n-1).
\end{eqnarray*}

\section{Linear operators in several variables}

The results of the preceding section extend readily to the case of
linear operators in $N$ variables $(x_{1},\ldots, x_{N})$. We
consider the vector space
\begin{equation}
{\mathcal{P}}_{n}^{N}=\{x_{1}^{i_{1}}\ldots
x_{N}^{i_{N}}\mid i_{1}+\cdots +i_{N}\leq n\},
\end{equation}
of polynomials of degree $n$ in $N$ variables, of dimension
$\binom{N+n}{n}$. We shall use the standard multi-index notation
whereby given a multi-index $I=(i_{1}\ldots i_{N})\in
{\mathbb{N}}^{N}$, we let
\begin{equation}
x^{\miI}:=x_{1}^{i_{1}}\ldots x_{N}^{i_{N}},\quad
D_{I}:=\frac{\partial}{\partial x_{1}^{i_{1}}}\cdots
\frac{\partial}{\partial x_{N}^{i_{n}}}.
\end{equation}
The notions of order, degree and deficiency are defined similarly
to the single-variable case. Thus, an operator
\begin{equation}
L=\sum_{|I|=0}^{r}a_{I}(x_{1},\ldots, x_{N})D_{I},
\end{equation}
will be of degree $d\in \mathbb{Z}$ if for almost all $I\in
\mathbb{N}^{N}$, there exists $c_{\miI\miJ}\neq 0 \in
(\mathbb{R}^{N})^2$ such that
\begin{equation}
  L[x^{\miI}]= \sum\limits_{|\miJ|=|\miI|+d}
  c_{\miI\miJ}\,x^{\miJ}\, .
\end{equation}
In order to define the deficiency, we fix again $n\in \mathbb{N}$,
and say that $L$ has {\em deficiency} $m \in \mathbb{Z}$ relative
to $\mathcal{P}_n^N$ if
\begin{equation}
L{\mathcal{P}}_{n}^N\subset {\mathcal{P}}_{n-m}^N,\;\,\mbox{but
}\, L{\mathcal{P}}_{n}^N\not\subset {\mathcal{P}}^N_{n-m-1}.
\end{equation}
As in the single variable case, it is easy to see that the
deficiency of an operator cannot exceed its order. For example,
the operator $D_{\miI}$ has order $|\miI|$, degree $-|\miI|$, and
deficiency $|\miI|$, and the operator $x^{\miI}$ has order zero,
degree $|\miI|$, and deficiency $-|\miI|$. Again, a more
significant example is obtained by introducing the Euler operator
\begin{equation}
E:=\sum_{i=1}^{n}x_{i}\frac{\partial}{\partial x_{i}},
\end{equation}
the Pochhammer operator
\begin{equation}
(\, n-E\,)_{k}:=(-1)^{k}(\, E-n \,)(\, E-(n-1)\,)\ldots(\,
E-(n-k+1)\,),
\end{equation}
and considering the operator
\begin{equation}
x^{\miI}(n-|\miJ|-E)_{k}D_{\miJ}.
\end{equation}
This operator has order $|\miJ|+k$, degree $|\miI|-|\miJ|$ and deficiency
$|\miJ|+k-|\miI|$. Again, if we let ${\linopspace}_{r,m}$ denote the
vector space of linear differential operators in $N$ variables
$(x_{1},\ldots, x_{N})$, with polynomial coefficients, of order
$r$ and deficiency $m$, then we have
\begin{equation}
\dim\,{\linopspace}_{r,m}=\binom{N+r-m}{r-m}\binom{N+r}{r},
\end{equation}
or equivalently
\begin{equation}
{\linopspace}_{r,m}={\mathcal{P}}_{r-m}\otimes{\linopspace}_{r,r}.
\end{equation}
This formula is in agreement with the result obtained in
~\cite{FK} for operators in two variables, in which it was proved
that
\begin{equation}
\dim\,{\linopspace}_{r,0}={\binom{2+r}{r}}^{2}.
\end{equation}
The proof given in ~\cite{FK} was less direct and required an
analysis of the syzygies defined by the primitive ideals
associated to the irreducible representations of ${\mathfrak
{sl}}(3,\mathbb{R})$.

 It is easy to see that in contrast with the
single variable case, the order, degree and deficiency are not
sufficient to characterize an operator uniquely up to a non-zero
factor. We have:
\begin{proposition}
A basis for the vector space of linear differential operators with
polynomial coefficients, of order $r$, degree $d$ and deficiency
$m$, in $N$ variables is given by
\begin{equation}
x^{\miI}(n-|\miJ|-E)_{k}D_{\miJ},\quad
|\miI|=r-m,\,|\miJ|=r-d-m,\,k=d+m.
\end{equation}
This vector space is thus of dimension
\begin{equation}
\binom{N+r-m-1}{r-m}\binom{N+r-m-d-1}{r-m-d}.
\end{equation}
\end{proposition}

\section{Non-linear operators}

Our objective in this section is to show that the results of the
two preceding sections can be applied to prove a structure theorem
for a class of non-linear differential operators admitting
invariant polynomial subspaces. We shall see that these results
complement the structure theorems for operators preserving
simplicial modules which were proved in ~\cite{KMO}.

In dealing with non-linear operators it is convenient to identify
differential operators with functions on jet space.  To that end,
let
\begin{equation*}
  \cJ^r(\Rset) = \Rset \times \Rset^{r+1}
\end{equation*}
denote the bundle of $r$-jets of smooth maps from $\Rset$ to
$\Rset$. The $r$-th prolongation of a smooth, real-valued function
$f(x)$ is a section of $\cJ^r$, namely
\begin{equation*}
\pr_r(f)=(f,\, Df,\, D^2\!f,\ldots ,D^r\! f)
\end{equation*}
Thus, the action of an operator on a function of $x$ is the same
thing as the composition of a function of the jet variables with
the prolongation:
\begin{equation*}
  T[f] = T\circ \pr_r(f).
\end{equation*}
We introduce the standard jet  coordinates $x,u_0=u,u_1=u_x,
u_2=u_{xx},\ldots ,u_r$ on $\cJ^r$ so that
\begin{equation*}
  D^j[f]= u_j\circ \pr_r(f).
\end{equation*}

Henceforth, we fix $n$. By Propositions \ref{prop:maxdef} and
\ref{prop:Trm}, all linear operators of order $r$ are expressed
uniquely as polynomial linear combinations of the maximal deficiency
operators $K_{rj}$.
Thus,  a linear operator $K_{rj}$ maps $\cP_n$ to $\cP_{n-r}$; a
quadratically non-linear $K_{ri} K_{rj}$ maps $\cP_n$ to $\cP_{2(n-r)}$, a
cubically nonlinear $K_{ri} K_{rj} K_{rk}$ maps $\cP_n$ to $\cP_{3(n-r)}$,
etc.
This implies:
\begin{proposition}\label{nonlinmap} Every   operator ( linear or
  non-linear ) of   order $r$ can be uniquely expressed as
\begin{equation}
\label{eq:nonlin}
T:=p(x)+\sum_{i} p_{i}(x) K_{ri}+\sum_{i\leq j}p_{ij}(x)K_{ri} K_{rj}
+\sum_{i\leq j\leq k}p_{ijk}(x)K_{ri} K_{rj} K_{rk}+\cdots,
\end{equation}
The operator in question will map ${\mathcal {P}}_{n}$ to
${\mathcal {P}}_{n}$ if and only if
\begin{equation}
\label{eq:nonlincondition}
p\in {\mathcal{P}}_{n},\quad p_{i}\in {\mathcal{P}}_{r},\quad
p_{ij}\in {\mathcal{P}}_{2r-n}, \quad p_{ijk}\in
{\mathcal{P}}_{3r-2n},\;\ldots
\end{equation}
\end{proposition}

The above proposition has the following obvious consequence:
\begin{corollary}
An operator of order $r< {{n}\over{2}}$, mapping ${\mathcal
{P}}_{n}$ to ${\mathcal {P}}_{n}$ is necessarily linear. If $r<
{{2n}\over{3}}$, then the operator will have at most quadratic
non-linearities.
\end{corollary}
Conversely, the above proposition can be used to bound the degree of
any polynomial space which can be left invariant by a non-linear
operator. We have, for example:
\begin{corollary}
A second-order operator will preserve
a polynomial space of degree at most four.
\end{corollary}

The extension of these results to multivariate differential
operators acting on simplicial polynomial modules is
straightforward. We are interested in writing down all non-linear
differential operators of order $r$  that preserve
\begin{equation}
{\mathcal{P}}_{n}=\{x_{1}^{i_{1}}\ldots
x_{N}^{i_{N}}\mid i_{1}+\cdots +i_{N}\leq n\}.
\end{equation}
We recall that for fixed order $r$ the maximum deficiency that an
operator can attain is $r$, which is achieved by any of the
following {\em maximum deficiency} operators:
\begin{eqnarray}
& &K_{\miJ} \in {\linopspace}_{r,r}, \implies K_{\miJ}
=(n-|\miJ|-E)_{r-|\miJ|}\,
D^{\miJ},\\
 & &\quad \miJ= \{j_1,\dots,j_N\mid j_{1}+\cdots +j_{N}\leq r\}.
\end{eqnarray}

The extension of Proposition \ref{nonlinmap} to the case of
multivariate differential operators and polynomial modules is
straightforward by just substituting the simple indices $i,j$ into
multi-indices $I,J$, each of which can assume $\dim {\mathcal
L}_{r,r} = \binom{N+r}{r}$ different values. Similarly, the two
Corollaries hold verbatim in the multivariate case.

\noindent\noindent\textbf{Example 2.}\label{ex2}
  Write down all second order operators with quadratic non-linearities
  that map $\mathcal P_4$ into itself.  A direct application of
  Proposition \ref{nonlinmap} allows to write:
  \begin{equation}
    T[u]= \sum_{0\leq i\leq j\leq 2} p_{ij}\, K_{2i} K_{2j}
  \end{equation}
  where $p_{ij}=p_{ji}$ are constants, and where
  \begin{align*}
    K_{20} &= 6 u_0 -3 x u_1  + \tfrac{1}{2}\,x^2u_2,\\
    K_{21} &= 3 u_1 - x u_2,\\
    K_{22} &= u_2
  \end{align*}
  The resulting quadratic combination is the following six parameter
  family of operators:
  \begin{eqnarray*}
    T &=&p_{00} u_2^2 +
    p_{01}\,(-xu_2^2+3u_1u_2)+p_{02}\,(x^2u_2^2 -6x
    u_1u_2+12u_2u_0)\\
    &&+\, p_{11}\,(x^2 u_2^2+9u_1^2-6xu_1u_2)\\
    &&+\,p_{12}\,(-x^3 u_2^2 + 9 x^2 u_1u_2 - 6x(2u_2 u_0+ 3
    u_1^2) +36u_1u_0))\\
    &&+\,p_{22}\,\left(\tfrac{1}{4}\,x^4u_2^2 -3x^3u_1u_2+3x^2(2u_2u_0+3u_1^2
      )-36xu_1u_0+36u_0^2\right)
  \end{eqnarray*}
  If we are interested in finding only autonomous non-linear equations
  we need to consider the translation-invariant subfamily of the above
  operators, i.e. operators where the variable $x$ does not appear in
  the coefficients.  Imposing this condition leads to
  \[p_{10}=p_{12}=p_{22}=0;\quad p_{11}+p_{20}=0,\]
  so that the only autonomous second order non-linear operators
  which preserve the space $\mathcal P_4$ are:
  \begin{equation}
    \{u_{xx}^2\,,\;4u_{xx}u-3u_x^2\}.
  \end{equation}
  The first of these two operators is easily seen to map $\mathcal
  P_4$ into itself. The second operator is more interesting. Each of
  the non-linear terms $u_x^2$ and $u_{xx}u$ transform a
  $4\supth$-degree polynomial into a $6\supth$-degree polynomial.
  However, the linear combination $4u_{xx}u-3u_x^2$  cancels the
  coefficients of degree $6$ and $5$, and hence defines a map from
  $\cP_4$ to itself (see also \cite{Ga} and \cite{Sv}).

We investigate further in the analysis of autonomous non-linear
operators with invariant polynomial subspaces in the following
section.

\section{The algebra of polynomially non-linear operators}

In this section we continue the study non-linear operators that are
polynomial in the function and its first $n$ derivatives.  First, let
us fix some notation.  Let
\begin{equation*}
  \opalgebra=\Rset[x,u_0,\dots,u_n]
\end{equation*}
denote the commutative algebra of non-linear operators that can be
expressed as polynomials over $\Rset$ in $x$ and the derivatives
$u_j$.  Multiplication in this algebra is by pointwise multiplication,
rather than operator composition. We grade this algebra by total
degree in the $u_j$ variables:
\begin{equation*}
  \opalgebra = \bigoplus_{l=0}^\infty \opalgebra_\ell,
\end{equation*}
where
\begin{equation*}
  \opalgebra_\ell = \spn\{ x^j\, u_{i_1} u_{i_2} \cdots u_{i_\ell} \mid
  0\leq j<\infty\}.
  \end{equation*}

We will refer to the integer $\ell$ as an operator's degree of
non-linearity.  Thus, $\opalgebra_1$ is the vector space of linear operators,
$\opalgebra_2$ the vector space of quadratically non-linear operators, etc.
The vector space $\opalgebra_0$ is the space of constant operators.  For
example, the operator $x$ maps all of $\cP_n$ to $x$.  Thus, $\opalgebra_0$
is a subalgebra of $\opalgebra$, which is isomorphic to the polynomial
algebra in the variable $x$.  All the other $\opalgebra_\ell,\; \ell\geq 1$
are merely subspaces of $\opalgebra$, not subalgebras.

We further grade each
$\opalgebra_\ell$ according to the following monomial weighting scheme:
\begin{equation}
  \label{eq:wtdef}
 \wt(u_i)=n-i,\qquad \wt(x)=1.
\end{equation}
Thus,
\begin{equation}
  \label{eq:Tlgrading}
  \opalgebra_\ell = \bigoplus_{k=0}^\infty \opalgebra_{\ell,k},
\end{equation}
where
\begin{equation}
  \opalgebra_{\ell,k} = \spn\{ x^j\, u_{i_1} u_{i_2} \cdots u_{i_\ell} \mid
  j+n\ell-\sum_{s=1}^\ell i_s =k\}
\end{equation}
is the subspace generated by monomials having weight $k$. We will
refer to the integer $n-k$ as the {\em monomial deficiency}. If
$M=x^j u_{i_1} u_{i_2}\dots u_{i_\ell}$ is an operator monomial
with
\begin{equation}
  \label{eq:wtM}
  k=\wt(M)=j+n\ell-\sum_{s=1}^\ell i_s,
\end{equation}
then, in accord with the above-introduced meaning of deficiency, $M$
maps $\cP_n$ into $\mathcal P_{k}$, but not into $\mathcal
P_{k-1}$.

In other words, every  operator $T\in \opalgebra$ admits the unique
decomposition
\begin{equation}
  \label{eq:Tansatz}
  T = \sum_{\ell,k} T_{\ell,k},
\end{equation}
where
\begin{equation}
  \label{eq:Tkl}
  T_{\ell,k}=
  \sum_{\substack{0\leq i_1\leq i_2\leq \dots\leq
      i_\ell\leq n\\[2pt]
      k=j+n\ell-(i_1+i_2+\cdots+i_\ell)}} C_{ji_1
    i_2\dots i_\ell} x^j u_{i_1} u_{i_2}\dots u_{i_\ell},
\end{equation}
and where the sum is taken over finitely many values of $\ell$ and
$k$. For generic values of the coefficients $C_{ji_1\dots
i_\ell}$, the operator $T_{\ell,k}$ has deficiency $n-k$.
However, for certain very specific values of the coefficients, the
actual deficiency is greater than the monomial deficiency.  This
is so because in the linear combination (\ref{eq:Tkl}) there might
occur some cancellations in the terms of highest degree. We see
for instance in Example 2 that relative to $\cP_4$, the operator
\begin{equation*}
  4u_{xx} u-3u_x^2 = 4 u_2 u_0 - 3 u_1^2
\end{equation*}
has monomial deficiency $4-k=4-8+2=-2$. However, the actual deficiency
of this operator is zero.

Next, we describe generators for $\opalgebra$ that will allow us to precisely
determine the deficiency of an operator.  Following Proposition
\ref{prop:Krjsum}, let us re-introduce the linear, $n\supth$-order
operators of maximal deficiency:
\begin{equation}\label{eq:vurel}
  v_j = K_{nj} = \sum_{i=0}^{n-j} (-1)^{i}\,  \frac{x^i}{i!}\,
  u_{i+j}\,\quad
  j=0,\ldots,n.
\end{equation}
Thus,
\begin{align}
  \label{eq:vndef}
  v_n &= u_n, \\
  \label{eq:vn-1def}
  v_{n-1} &= u_{n-1} - x u_n,\\ \nonumber
  v_{n-2} &= u_{n-2} - x u_{n-1} + \tfrac{1}{2}\, x^2 u_{n-2},\\ \nonumber
  v_{n-3} &= u_{n-3} - x u_{n-2} + \tfrac{1}{2}\, x^2
  u_{n-3}-\tfrac{1}{6}\, x^3
  u_{n-4},  \\ \nonumber
  \vdots
\end{align}
These operators transform all elements of $\mathcal P_n$ into a
constant.  In other words, the $v_j$ are the operator duals \cite{KMO}
to the monomial basis of $\mathcal P_n$:
\begin{equation*}
  v_j[x^k/k!] = \delta_j{}^k,\quad k=0,1,\dots,n.
\end{equation*}
More, generally let us write
\begin{equation}
  \label{eq:tudef}
  \tu_j(t) =  \sum_{i=0}^{n-j} \,
  u_{i+j}\,\frac{t^i}{i!}\quad
  j=0,\ldots,n.
\end{equation}
In this way,
\begin{equation*}
  v_j =\tu_j(-x).
\end{equation*}
\begin{proposition}
  \label{prop:tu1param}
  We have
  \begin{equation}
    \label{eq:tu1param}
    \tu_j(s+t) =  \sum_{i=0}^{n-j} \tu_{i+j}(s)\,
    \frac{t^i}{i!}.
  \end{equation}
\end{proposition}
\noindent\textbf{Proof}
  We have
  \begin{align*}
    &\tu_j{}'(t) = \tu_{j+1}(t),\qquad j=0,1,\dots,n-1,\\
    &\tu_n{}'(t) = 0, \\
    &\tu_j(0) = u_j.
  \end{align*}
  Hence, $u_j\mapsto \tu_j(t)$ defines a 1-parameter transformation
  group of $\opalgebra$.  The desired result follows immediately.
  \qed

\noindent Now, we can invert the relations \eqref{eq:vurel}, and
express the $u_j$ in terms of the $v_j$.
\begin{proposition}
  \label{prop:uvrel}
  For $j=0,\dots, n$, we have
  \begin{equation}
    \label{eq:uvrel}
    u_j = \sum_{i=0}^{n-j}  v_{i+j}\, \frac{x^i}{i\,!},
  \end{equation}
\end{proposition}
\noindent\textbf{Proof}
  We apply \eqref{eq:tu1param} with
  $s=-x$ and $t=x$. \qed

Proposition \ref{prop:uvrel} shows that $x, v_0,\dots, v_n$ freely
generate the algebra $\opalgebra$.  The relations \eqref{eq:vurel} and
\eqref{eq:uvrel} are homogeneous relative to the weights
\eqref{eq:wtdef}.  Hence, setting
\begin{equation}
  \label{eq:vwtdef}
  \wt(v_j) = n-j,
\end{equation}
we recover the grading by monomial deficiency relative to this basis.
We now deepen the grading by defining
\begin{equation*}
  \opalgebra_{\ell,k,m} = \spn\{ x^m\, v_{i_1} v_{i_2} \cdots v_{i_\ell} \mid
  m+n\ell-\sum_{s=1}^\ell i_s =k\},
\end{equation*}
so that
\begin{equation*}
  \opalgebra_{\ell,k} = \bigoplus_{m=0}^k \opalgebra_{\ell,k,m}.
\end{equation*}

\begin{proposition}
  \label{prop:mlkgrading}
  The elements of $\opalgebra_{\ell,k,m}$ have monomial deficiency $n-k$ and
  actual deficiency $n-m$.   Consequently, every operator $T\in \mathbb{R}[x,u_0,\ldots,u_n]$ has
deficiency $n-m$, where $m$ is the $x$-degree of the polynomial
  \begin{equation*}
    T=Q(x,v_0,\dots, v_n)
  \end{equation*}
  that expresses $T$ relative to the $x, v_j$ basis.
\end{proposition}
\noindent\textbf{Proof}
  Since the relations \eqref{eq:vurel} \eqref{eq:uvrel} are
  homogeneous relative to the weighting scheme \eqref{eq:wtdef}
  \eqref{eq:vwtdef}, the elements of $\cT_{\ell,k,m}$ have weight
  $k$, and hence have monomial deficiency $n-k$.  Since the operators
  $v_j$ map all polynomial to constants, an operator $Q(x,v_0,\dots,
  v_n)\in \opalgebra$, having $m$ as its $x$-degree, maps $\cP_n$ to $\cP_m$,
  but not to $\cP_{m-1}$. Therefore such an operator has deficiency
  $n-m$. \qed

The key application of the above grading has to do with the
decomposition of an operator according to monomial deficiency. Our
analysis would be greatly simplified if we could be certain that
the decomposition of a non-linear operator $T$ according to
monomial deficiency respects the actual deficiency.  In other
words, when considering operators of a fixed deficiency, no
generality is lost by considering operators that are homogeneous
in degree of non-linearity and monomial deficiency.
\begin{corollary}
  \label{cor:decompdeficiency}
  Let $T$ be a non-linear operator whose deficiency is $n-m$ or more,
  i.e., $T$ maps $\cP_n$ into $\cP_{m}$.  Let $T_{\ell,k}$ be the
  summands of the decomposition of $T$ according to degree of
  non-linearity and monomial defiency as per \eqref{eq:Tansatz}
  \eqref{eq:Tkl}.  Then, each $T_{\ell,k}$ also maps $\cP_n$ into
  $\cP_m$.
\end{corollary}
\section{Autonomous, non-linear operators}
Our main focus in this section is the subalgebra
\begin{equation}
  \cA=\Rset[u_0,\cdots, u_n]\subset \opalgebra
\end{equation}
of translation-invariant non-linear opearators.
The subalgebra inherits the bi-grading relative to degree of
non-linearity and monomial deficiency, with
\begin{align}
  &\cA=\bigoplus_{\ell=0}^\infty \bigoplus_{k=0}^{n\ell}
  \cA_{\ell,k},\intertext{where}
  &\cA_{\ell,k} = \cA\cap \opalgebra_{\ell,k} =
  \spn\{ u_{i_1} u_{i_2} \cdots u_{i_\ell} \mid
  n\ell-\sum_{s=1}^\ell i_s =k\}
\end{align}
Our key result in this section is the characterization of the
deficiency of autonomous operators.  In other words, we describe
$\cA\cap \opalgebra_{\ell,k,m}$.

The obvious approach to construct non-linear operators of
deficiency $m$ would be to write a generic polynomial $p(x)\in
\mathcal P_n$ with indeterminate coefficients, act on it by
(\ref{eq:Tansatz}) where the degree of the possible
non-linearities is bounded by Proposition \ref{nonlinmap} and
impose that the coefficients of all the terms in $x^j$ for $j>n-m$
vanish. However, based on the useful concept of deficiency we
choose here to adopt a somewhat different approach.

We seek operators that are both
translation-invariant and that have maximum deficiency.  To this end,
we define
\begin{align}
    \label{eq:xidef}
   &\xi =\frac{u_{n-1}}{u_n}; \\
   \label{eq:Iurel}
   &I_{n-j} = \tu_j(-\xi) = \sum_{i=0}^{n-j} (-1)^{k}
    \,u_{i+j} \frac{\xi^i}{i!}   ,\qquad
   j=0,\ldots,n,
\end{align}
and note that
\begin{equation}
  \label{eq:I0I1}
  I_0 = u_n,\qquad I_{1} = 0.
\end{equation}
The jet space function $\xi$ is only defined on open neighborhood
$u_n\neq 0$ of $\cJ^n$.  Thus, the operators $I_j$ are defined
only for elements of $\cP_n^{\times}= \cP_n\setminus \cP_{n-1}$,
the set of polynomials of degree exactly equal to $n$.  We can
still speak of the deficiency of such operator, but this has to be
understand in terms of $\cP^{\times}_n$ rather than $\cP_n$.
\begin{proposition}\label{prop:Ik}
  The non-linear operators $I_2,I_3\dots, I_{n}$ are
  translation-invariant, and have maximum deficiency $n$.  In other
  words, these autonomous, nonlinear operators transform every
  $n\supth$ degree polynomial into a constant.
\end{proposition}
\noindent\textbf{Proof}
  By definition,
  \begin{equation*}
    v_j = \tu_j(-x),\qquad
    I_{n-j} = \tu_j(-\xi).
  \end{equation*}
  As well,
  \begin{equation}
    \label{eq:xixrel}
    \xi=x+\frac{v_{n-1}}{v_n}.
  \end{equation}
  Hence, Proposition \ref{prop:tu1param} implies  that in addition to
  \eqref{eq:Iurel} we also have
  \begin{equation}
    \label{eq:Ivrel}
    I_{n-j} =\sum_{k=0}^{n-j} (-1)^{k}\frac{1 }{k!}
    \,v_{k+j}\left( \frac{v_{n-1}}{v_n}\right)^k .
  \end{equation}
  Hence, the operators $I_j$ are polynomials of the $v_j$ divided by a
  certain power of $v_n=u_n$.  Therefore, these operators are both
  translation invariant and of maximal deficiency. \qed

\noindent We can also invert the relations \eqref{eq:Iurel}, and
express the $u_j$ in terms of the $I_j$.
\begin{proposition}
  \label{prop:uIrel}
  For $j=0,\dots, n$, we have
  \begin{align}
    \label{eq:uIrel}
    u_j &= \sum_{i=0}^{n-j}  I_{n-i-j}\, \frac{\xi^i}{i\,!},\\
    \label{eq:uIrel1}
       &= \sum_{i=0}^{n-j}  I_i\, \frac{\xi^{n-j-i}}{(n-j-i)\,!},
  \end{align}
\end{proposition}
\noindent\textbf{Proof}
  We apply \eqref{eq:tu1param} with
  $s=-\xi$ and $t=\xi$. \qed

\noindent Thus, relations \eqref{eq:Iurel} \eqref{eq:uIrel} tell
us that the operators $I_0,\xi,I_2\dots,I_{n}$ also generate the
algebra of autonomous operators. These relations are homogeneous
with respect to the monomial weights defined in \eqref{eq:wtdef}.
Hence, setting
\begin{equation}
  \label{eq:Iwtdef}
  \wt(I_j) = j,\qquad \wt(\xi)=1
\end{equation}
we recover the grading by monomial deficiency relative to this basis.

Unfortunately, the operators $I_j, \xi$ are not polynomials in the
$u_j$, and hence are \emph{not} elements of $\cA$.  Let us
therefore consider the larger algebra
\begin{equation}
  \label{eq:hAdef}
  \hcA= \Rset[I_0,\xi,I_2,I_3,\dots,I_n]
\end{equation}
generated by $\xi$ and the autonomous operators of maximal
deficiency.  Thanks to \eqref{eq:uIrel} we know that $\cA\subset
\hcA$, but this inclusion is strict.

We now deepen the grading of $\hcA$ by defining
\begin{equation*}
  \hcA_{\ell,k,m} = \spn\{ \xi^m\, I_{i_1} \cdots I_{i_\ell} \mid
  m+i_1+\cdots + i_\ell =k\},
\end{equation*}
so that
\begin{equation*}
  \hcA = \bigoplus_{\ell=0}^\infty \bigoplus_{k=0}^{n\ell}  \bigoplus_{m=0}^k
  \hcA_{\ell,k,m}.
\end{equation*}

\begin{proposition}
  \label{prop:autmlkgrading}
  We have
  $\hcA_{\ell,k,m} = \hcA\cap \opalgebra_{\ell,k,m}$,
  i.e., the elements of $\hcA_{\ell,k,m}$ have monomial deficiency $n-k$ and
  actual deficiency $n-m$.
\end{proposition}
\noindent\textbf{Proof}
  This follows  from \eqref{eq:xixrel} \eqref{eq:Ivrel}, and the fact
  that the operators $v_j$ have maximal deficiency. \qed

\noindent In other words, the deficiency of an autonomous operator is
$n$ minus the $\xi$-degree of the polynomial that expresses that
operator relative to the $\xi, I_j$ basis. We are left with the
question of the nature of the inclusion of $\cA$ in $\hcA$.  In other
words, which polynomials in $\xi, I_j$ define true polynomial
operators.
\begin{theorem}
  \label{thm:autopchar}
  Let
  $T=P(x,u_0,\dots,u_n)\in\opalgebra$ be a non-linear operator, and let
  $T=Q(x,v_0,\dots, v_n)$ be
  the  expression of this operator
  relative to the
  non-autonomous generators $x, v_j$.
  Then $T$ is autonomous, i.e. $P$ is independent of the variable $x$
  if and only if
  \begin{equation*}
    T=Q(\xi,I_n,\dots,I_2,0,I_0).
  \end{equation*}
  In this case, the deficiency of $T$ is equal to $n$ minus the
  $\xi$-degree of the polynomial $Q(\xi,I_n,\dots, I_0)$.
\end{theorem}
\noindent\textbf{Proof}
  By Propositions \ref{prop:uvrel} and \ref{prop:uIrel}, we have
  \begin{equation*}
    u_j = \sum_{i=0}^{n-j}  v_{i+j}\, \frac{x^i}{i\,!} =
    \sum_{i=0}^{n-j}  I_{n-i-j}\, \frac{\xi^i}{i\,!}, \quad j=0,1,\dots,
    n,
  \end{equation*}
  where, as we noted before, $I_{1}=0$, and $I_0 = v_n = u_n$.
  \qed

\noindent\textbf{Example 3.}
  \label{ex3}
  Let us recast the analysis began in Example 2 in terms of
  the above operator bases.  Relations \eqref{eq:uvrel} and
  \eqref{eq:uIrel} take the form
  \begin{align*}
    u_4 &= v_4 &&= I_0;\\
    u_3 &= v_3 + x v_4 &&= \xi I_0 ,\\
    u_2 &= v_2 + x v_3 + \tfrac{1}{2}\, x^2 v_4 &&= I_2 +
    \tfrac{1}{2}\,  \xi^2\, I_0\;\\
    u_1 &= v_1 + x v_2 +  \tfrac{1}{2}\, x^2 v_3 + \tfrac{1}{6}\, x^3
    v_4  && = I_3 + \xi I_2 +  \tfrac{1}{6}\, \xi^3 I_0;\\
    u_0 &= v_0 + x v_1 +  \tfrac{1}{2}\, x^2 v_2 + \tfrac{1}{6}\, x^3
    v_3+ \tfrac{1}{24}\, x^4 v_4  && = I_4 + \xi I_3 +  \tfrac{1}{2}\,
    \xi^2 I_2 + \tfrac{1}{24}\, \xi^4 I_0.
  \end{align*}
  Our goal is to write down all autonomous operators with quadratic
  non-linearity and zero deficiency, i.e., operators that map
  $\cP_4$ into itself.  By Proposition \ref{prop:autmlkgrading} and the
  above relations, we are obliged to consider polynomials that are
  quadratic in $I_0, I_1, I_2,I_4$ and that have degree $4$ or less in the
  $\xi$ variable.  The question is: which operators of such form are
  quadratic in $u_0,u_1,u_2, u_3,u_4$?  By Corollary
  \ref{cor:decompdeficiency}, no generality is lost by considering
  operators of a fixed monomial deficiency.  Evidently, if the
  monomial deficiency is $0$ or more, the operator will preserve
  $\cP_4$.  Hence, we must consider operators having monomial
  deficiency $-4,-3,-2,-1$, which corresponds to $k=8,7,6,5$,
  respectively.  The most general operator having $k=8$ is a multiple
  of the monomial $(u_0)^2$.  Such an operator will have actual
  deficiency of $-4$ as well.  Similarly reasoning holds for operators
  with $k=7$; these are multiples of $u_0 u_1$.  Let us consider the
  case $k=6$.  The ansatz is now
  \begin{align*}
    C_{02}\, u_0 u_2 + C_{11}\, u_1^2 &= \frac{1}{144}(3 C_{02} +
    4 C_{11})\,v_4^2 \,x^6 + \frac{1}{24}(3C_{02} + 4 C_{11})\,
    v_3 v_4\, x^5 + \mbox{l.o.t.} \\
    &= \frac{1}{144}(3 C_{02} +
    4 C_{11})\,I_0^2 \,\xi^6 +0\, \xi^5+ \mbox{l.o.t.} \\
  \end{align*}
  Hence, such an operator preserves $\cP_4$ if and only if is a
  multiple of $4 u_0 u_2- 3 u_1^2$, as has already been proven by
  other methods.

  Finally, let us consider the case $k=5$.  Now, the ansatz is
  \begin{align*}
    C_{03}\, u_0 u_3 + C_{12}\, u_1u_2 &= \frac{1}{24}(C_{03} + 2
    C_{12})\,v_4^2 \,x^5 + \frac{5}{24}(C_{03} + 2 C_{12})\, v_3 v_4
     x^4 + \mbox{l.o.t.} \\
    &= \frac{1}{24}(C_{03} + 2 C_{12})\,I_0^2 \,\xi^5 +
    0\, \xi^4 + \mbox{l.o.t.}
  \end{align*}
  Hence, such an operator preserves $\cP_4$ if and only if is a
  multiple of $2 u_0 u_3- u_1 u_2$.  Indeed, because $I_1=0$, the
  above calculation shows that this operator has deficiency $1$,
  i.e. it maps $\cP_4$ into $\cP_3$.

\subsection*{Autonomous operators with quadratic
non-linearities}

We restrict from here on to $\opalgebra_2$, the vector space of
operators with homogeneously quadratic non-linearity.  We present
a complete characterization of such operators in terms of
deficiency, thereby extending the results of Svirshchevskii
\cite{Sv} and Galaktionov \cite{Ga}. Our analysis can be extended
to operators with higher non-linearities, but this shall be
treated elsewhere.

Following \eqref{eq:Tlgrading}, we let
\begin{equation*}
  \cQ= \bigoplus_{k=-n}^{n} \cQ_{k}
\end{equation*}
be the linear space of quadratic autnomous non-linear operators up
to order $n$, graded according to monomial deficiency.
Thus,
\begin{eqnarray}\label{eq:Qkdef}
  & & \cQ=\cA_2 = \spn\{ u_i u_j \mid 0\leq i \leq j \leq n \},\\
  && \dim \cQ = \binom{n+2}{2},\\
  & &\cQ_k= \cA_{2,n-k}=\spn \{ u_{i} u_{j} \mid i+j=n+k,\quad 0\leq
  i,j\leq n\},\\
  & &\dim \cQ_{k} =\left\lfloor
    \frac{n-\vert k\vert}{2}\right\rfloor+1,
\end{eqnarray}
where
and $\lfloor \cdot
\rfloor$ denotes the floor function.

Each of the elements of the above basis of $\cQ_{k}$ has a
different order $r$, the minimum and maximum orders for each $k$
being:
\begin{equation}\label{eq:rminmax}
\rkmin =\left\lceil\frac{n+k}{2}\right\rceil,\qquad
\rkmax= \min(n,n+k)
\end{equation}
All of the above basis elements have the same deficiency $k$, i.e.
they map $\mathcal P_n$ into $\mathcal P_{n-k}$.  However, Example
2 and Example 3 show that within each $\cQ_k$ there can be
elements whose deficiency is greater than $k$.

Remarkably, each $\cQ_k$ possesses an adapted basis that further
grades it according to operator order and deficiency.  In this regard,
for $\rkmin\leq r\leq \rkmax$, let us introduce the operators
\begin{equation}
  \label{eq:Qkrdef}
  Q_{k,r} =
  \sum_{i=k+n-r}^r\!\!\!\! (-1)^i \binom{i-k+1}{n-r+1}
    \binom{n-i}{n-r} \,u_i\, u_{n+k-i},
\end{equation}
We are now ready to state the main result of this Section:
\begin{theorem}
  \label{thm:Qkr}
  The operator $Q_{k,r}$ has monomial deficiency $k$, order $r$, and
  deficiency
  \begin{equation}
    \label{eq:mkrdef}
    m(k,r)= k+ 2\left(r-\rkmin\right).
  \end{equation}
  Furthermore, $\{ Q_{k,r} \mid \rkmin \leq r \leq \rkmax\}$ forms a
  basis of $\cQ_{k}$.
\end{theorem}
This result is similar to that proved in Proposition
\ref{prop:uniquelinear} for linear operators: up to a scalar multiple
there is only one quadratic autonomous operator with a given order
$r$, monomial deficiency $k$, and deficiency $m(k,r)$.  In Appendix A
we show the explicit form of these operators for $n=4,5$ and $6$.

We will prove the theorem by finding a generating function for the
operators $Q_{k,r}$.  This requires some new notation.
For a bivariate formal series,
\begin{equation*}
  p(z,w) = \sum_{i,j\geq 0} p_{ij} \,z^i w^j,
\end{equation*}
let us define
\begin{equation}
  \label{eq:gpxydef}
  \cG\{p(z,w)\}=\sum_{i\geq 0} p_{i,i} z^i
\end{equation}
to be the series formed from terms where the two variables have equal
exponents.
We will also adopt the convention that
\begin{equation}
  \binom{a}{i}  =
  \begin{cases}
    \displaystyle \frac{a(a-1)\cdots (a-i+1)}{i!}, & i\geq 0\\
    0 & i<0
    \end{cases}
\end{equation}


\noindent\textbf{Proof}[ Proof of Theorem \ref{thm:Qkr} ]
  We extend the
  definition \eqref{eq:Qkrdef} of $Q_{k,r}$ to all  $0\leq
  r\leq n$ by setting
  \begin{equation}
    \label{eq:Qkrdef1}
    Q_{k,r} = \sum_{i=\max(0,k)}^{\min(n,n+k)}\!\!\! (-1)^i
    \binom{i-k+1}{n-r+1}
    \binom{n-i}{n-r} \,u_i\, u_{n+k-i}.
  \end{equation}
  We can now form a generating function for $Q_{k,r}$ as follows:
  {\small
  \begin{align}
    \label{eq:th11}
    Q(z,t)&= \sum_{k=-n}^{n} \sum_{r=0}^n Q_{k,r}\, z^{n-r}\, t^{n-k}\\
    \nonumber
    &= \sum_{i,j=0}^n \sum_{r=0}^n (-1)^i
    \binom{n-i}{n-r} \binom{n-j+1}{n-r+1} u_i u_j\, z^{n-r} t^{2n-i-j} \\
    \nonumber &= \sum_{i,j=0}^n (-1)^i
    z^{-1}\cG\{z(1+z)^{n-i} (1+w)^{n-j+1}\} u_i u_j \,t^{2n-i-j}\\
    \label{eq:th12}
    &= \sum_{i,j=0}^n \sum_{p,q=0}^{n}(-1)^i
    \binom{s}{n-i-p} z^{-1}\cG\{z(1+z)^{n-i} (1+w)^{n-j+1}\} I_p I_q
    \frac{\xi^s}{s!}\,t^{2n-i-j},
  \end{align}}\par\noindent
  where $s=2n-i-j-p-q$,
  and where we used relation \eqref{eq:uIrel1} to replace the $u_j$
  with the $I_j$, the autonomous generators of maximal deficiency
  defined by \eqref{eq:Iurel}.
  Let us extend the definition of $I_j$
  to all $j\geq 0$ by setting
  \begin{equation}
    \label{eq:iurelext}
    I_j = \sum_{i=0}^j(-1)^i u_{n+i-j} \frac{\xi^i}{i!},
  \end{equation}
  and agreeing that $u_j=0$ for $j<0$.  However, we must
  note that the newly defined autonomous operators $I_j,\; j>n$ are no
  longer maximal deficiency operators.  Interchanging the summation
  order in \eqref{eq:th12} and re-indexing with
  \begin{equation*}
    i\to n-p-i,\qquad j \to n-q-j
  \end{equation*}
  we obtain
  {\small
  \begin{align}
    \nonumber
    Q(z,t) &=  \sum_{p,q=0}^{n}\sum_{i=0}^{n-p} \sum_{j=0}^{n-q}
    (-1)^i \binom{s}{n-p-i}
    z^{-1}\cG\{z(1+z)^{n-i} (1+w)^{n-j+1}\} I_p I_q
    \frac{\xi^s}{s!}\,t^{2n-i-j}
    \\ \nonumber
    &=  \sum_{p,q=0}^{n}\sum_{i=0}^{n-p} \sum_{j=0}^{n-q}
    (-1)^i \binom{s}{i}
    z^{-1}\cG\{z(1+z)^{p+i} (1+w)^{q+j+1}\} I_p I_q
    \frac{\xi^{s}}{s!}\,t^{p+q+s},
  \end{align}}\par
  \noindent
  with $s=i+j$ henceforth. Relation \eqref{eq:uIrel1} holds
  for $j<0$, thanks to the extended definition \eqref{eq:iurelext} of
  $I_j,\; j>n$. Hence,
  {\small
    \begin{align}\nonumber
      Q(z,t)&=\sum_{p,q=0}^{2n}\sum_{i=0}^{2n-p-q} \sum_{j=0}^{2n-p-q-i}
      (-1)^i \binom{s}{i}
      z^{-1}\cG\{z(1+z)^{p+i} (1+w)^{q+j+1}\} I_p I_q
      \frac{\xi^{s}}{s!}\,t^{p+q+s}\\
      \nonumber
      &= \sum_{p,q=0}^{2n}\sum_{s=0}^{2n-p-q} \sum_{i=0}^{s}
      (-1)^{n-p-i} \binom{s}{i}
      z^{-1}\cG\{z(1+z)^{p+i} (1+w)^{q+1+s-i}\} I_p I_q
      \frac{\xi^{s}}{s!}\,t^{p+q+s}\\
      \label{eq:th13}
      &= \sum_{p,q=0}^{2n}\!\!\sum_{s=0}^{2n-p-q} \!\!
      (-1)^{n-p}  z^{-1}\cG\{z(1+z)^{p} (1+w)^{q+1}(w-z)^s\} I_p I_q
      \frac{\xi^{s}}{s!}\,t^{p+q+s}
    \end{align}}
  For every positive integer $s$, let us introduce the generating
  function
  \begin{equation}
    \label{eq:phisdef}
    \phi(p,q,s;z) = \cG\{(1+z)^p(1+w)^q (w-z)^s\} = \sum_{\rho=0}^\infty
    \phi_{p,q,\rho,s} \,z^\rho,
  \end{equation}
  where
  \begin{equation}
    \label{eq:phisdef2}
    \phi_{p,q,\rho,s} = \sum_{i=0}^\rho (-1)^{\rho-i} \binom{p}{i}
    \binom{q}{2\rho-s-i} \binom{s}{\rho-i} ,
  \end{equation}
  Thus,
  \begin{align}
    \label{eq:phiprop3}
    \phi_{p,q,\rho,s} &=  0 ,\qquad s>2\rho,\\ \nonumber
    \phi_{p,q,\rho,2\rho} &= (-1)^\rho  \binom{2\rho}{\rho},\\ \nonumber
    \phi_{p,q,\rho,2\rho-1} &=  (q-p)\binom{2\rho-1}{\rho},\\ \nonumber
    \phi_{p,q,\rho,2\rho-2} &= \left(\binom{q}{2}-
      \binom{p}{2}\right)\binom{2\rho-2}{\rho} -
    p\,q\binom{2\rho-2}{\rho-1}
  \end{align}
  Before continuing, let us note the following properties of this function:
  \begin{gather}
    \label{eq:phiprop1}
    \phi(p,q,s;z) = (-1)^s \phi(q,p,s;z)\\
    \label{eq:phiprop2}
    \phi(p,q,s+1,z) = \phi(p,q+1,s;z) - \phi(p+1,q,s;z)
  \end{gather}
  This function is relevant to our proof, because
  \begin{equation}
    \cG\{z(1+z)^{p} (1+w)^{q+1}(w-z)^s\} =
    \phi(p+1,q+1,s;z)-\phi(p,q+1,s;z).
  \end{equation}
  Hence,
  \begin{equation}
    Q(z,t) = \sum_{k=-n}^n \sum_{r=0}^n \sum_{s=0}^{2n}
    Q_{k,r,s} \frac{\xi^s}{s!} \,z^{n-r} \,t^{n-k}
  \end{equation}
  where
  {\small
    \begin{gather}
      \label{eq:Qkrdef2}
      Q_{k,r,s}  =Q_{k,n-\rho,s}= \!\!\!\sum_{p+q=n-k-s}\!\!\!
      (-1)^{n-p} (\phi_{p+1,q+1,\rho+1,s}-\phi_{p,q+1,\rho+1,s})\,
      I_p I_q,\\
      \label{eq:Qkrdef3}
      \sum_{r=0}^n Q_{k,r,s} z^{n-r} = \!\!\!\sum_{p+q=n-k-s}\!\!\!
      (-1)^{n-p} z^{-1}(\phi(p+1,q+1,s;z)-\phi(p,q+1,s;z))\, I_p I_q.
    \end{gather}}
  \par\noindent Let $m(k,r)$ denote the deficiency of $Q_{k,r}$. By
  Proposition \ref{thm:autopchar}, $m(k,r)$ is equal to $n$ minus the
  largest value of $s$ for which $Q_{k,r,s}\neq 0$.  By
  \eqref{eq:phiprop3} and \eqref{eq:Qkrdef2}, we know that
  $Q_{k,n-\rho,s}=0$ for $s>2\rho+2$.  For $s=2\rho+2$, we have
  \begin{equation*}
    \phi_{p+1,q+1,\rho+1,2\rho+2} - \phi_{p,q+1,\rho+1,2\rho+2}
    =(-1)^{\rho+1}
    \left(\binom{2\rho+2}{\rho+1}-\binom{2\rho+2}{\rho+1}\right)=0,
  \end{equation*}
  and hence, $Q_{k,n-\rho,s}=0$ for $s=2\rho+2$.

  The analysis now breaks up into two cases.  Suppose that $n-k$ is
  even. For $s=2\rho+1$ we again use \eqref{eq:phiprop3} and
  \eqref{eq:Qkrdef2} to obtain
  \begin{equation*}
      Q_{k,n-\rho,2\rho+1}
      =\sum_{p+q=n-k-2\rho-1} \!\!\!\!\!
      (-1)^{n-p+\rho+1}\binom{2\rho+1}{\rho}  \, I_p
      I_q =0,
  \end{equation*}
  because of $p,q$ symmetry. For $s=2\rho$, we have
  \begin{equation*}
      Q_{k,n-\rho,2\rho}
      =\sum_{p+q=n-k-2\rho} \!\!\!\!\!
      (-1)^{n-p-\rho}\, \frac{(2r+p+q+2)}{2(\rho+1)}\,
        \binom{2\rho}{\rho}   \, I_p
      I_q \neq 0.
  \end{equation*}
  Therefore, if $n-k$ is even, we have $m(k,r) = n-2\rho = 2r-n$.  This
  agrees with \eqref{eq:mkrdef}.

  Next, suppose that $n-k$ is odd.  Now for $s=2\rho+1$ we have
   \begin{equation*}
      Q_{k,n-\rho,2\rho+1}
      =\sum_{p+q=n-k-2\rho-1} \!\!\!\!\!
      (-1)^{n-p+\rho+1}\binom{2\rho+1}{\rho}  \, I_p
      I_q \neq 0,
  \end{equation*}
  because now $p$ and $q$ have the same parity.  Therefore, if $n-k$
  is odd, then $m(k,r)=n-2\rho-1=2r-n-1$.   Again, this agrees with
  \eqref{eq:mkrdef}. \qed

\section{Separation of variables in non-linear evolution equations}

One of the main applications of the results of this paper lies in
the method of  separation of variables in non-linear evolution
equations. Let $u\in\mathcal P_n$ and $T$ be a non-linear
differential operator with  deficiency $m\geq 0$. This means that
$T\mathcal P_n\subset\mathcal P_n$, or more explicitly
\begin{equation}
T\left[\sum_{i=0}^n p_i x^i\right]= \sum_{i=0}^n
f_i(p_1,\dots,p_n) x^i
\end{equation}
The non-linear evolution equation
\begin{equation}\label{eq:evolution}
u_{t}=T[u]=p(x,u,u_1,\dots,u_n)
\end{equation}
admits separable solutions of the form
\begin{equation}
u(x,t)=\sum_{i=0}^{n}\varphi_{i}(t)\,x^{i},
\end{equation}
where the functions $\varphi(t)$ satisfy the following system of
first order ordinary differential equations:
\begin{equation}
\dot \varphi_i = f_i(\varphi_1,\dots,\varphi_n),\qquad
i=0,\dots,n.
\end{equation}
These results extend immediately to the multivariate case. This
technique has been used by King to find new exact multidimensional
solutions of non-linear diffusion equations \cite{K1,K3} and exact
solutions of high order thin film equations \cite{K2}. In a more
mathematical context, Galaktionov \cite{Ga} and Svirshchevskii
\cite{Sv} have analyzed non-linear operators that preserve low
dimensional spaces spanned by polynomials and trigonometric
functions.

\noindent\textbf{Example 4} Consider the following non-linear
evolution equation for $u=u(t,x)$:
\begin{equation}
u_t=7\left(u\,u_{xxxx} - \frac{5}{2}\, u_x\,u_{xxx} +
\frac{45}{28}\, u_{xx}^2\right)
\end{equation}
This equation has solutions in $\mathcal P_8$ of the form
\begin{equation}
u(x,t)=\sum_{i=0}^{8}\varphi_{i}(t)\,x^{i},
\end{equation}
where the functions $\varphi_{i}(t)$ satisfy the following first
order system:
\begin{eqnarray*}
\dot\varphi_0 &=& 45\,\varphi_2^2 - 105\,\varphi_1\varphi_3+168\,\varphi_0\varphi_4 \\
\dot\varphi_1 &=& 60\,\varphi_2\varphi_3-252\,\varphi_1\varphi_4+840\,\varphi_0\varphi_5\\
\dot\varphi_2 &=& 90\,\varphi_3^3-132\,\varphi_2\varphi_4-210\,\varphi_1\varphi_5+2520\,\varphi_0\varphi_6\\
\dot\varphi_3 &=& 108\,\varphi_3\varphi_4-360\,\varphi_2\varphi_5+420\,\varphi_1\varphi_6+5880\,\varphi_0\varphi_7\\
\dot\varphi_4 &=& 108 \,\varphi_4^2-135\, \varphi_3\varphi_5-330\,\varphi_2\varphi_6+2205\,\varphi_1\varphi_7+11760\, \varphi_0\varphi_8\\
\dot\varphi_5 &=& 108\,\varphi_4\varphi_5-360\,\varphi_3\varphi_6+420\,\varphi_2\varphi_7+5880\,\varphi_1\varphi_8\\
\dot\varphi_6 &=& 90\,\varphi_5^3-132\,\varphi_4\varphi_6-210\,\varphi_3\varphi_7+2520\,\varphi_2\varphi_8\\
\dot\varphi_7 &=& 60\,\varphi_5\varphi_6-252\,\varphi_4\varphi_7+840\,\varphi_3\varphi_8\\
\dot\varphi_8 &=& 45\,\varphi_6^2
-105\,\varphi_5\varphi_7+168\,\varphi_4\varphi_8
\end{eqnarray*}

%
%

\section*{Appendix A}

In this Appendix we list the specialized autonomous operator basis
$Q_{k,r}$ described in Theorem 1.  Recall that each $Q_{k,r}$ has
monomial deficiency $k$, order $r$, and deficiency $m=m(k,r)$.  The following
tables for $n=4,5,6$ display the non-linear operators with $\{k,r,m\}$
in the ranges
\[
-n\leq\,k\,\leq n,\qquad \left\lceil\frac{n+k}{2}\right\rceil\leq
\,r\,\leq \min(n,n+k),\qquad m= k+ 2(r-r_{min})\,.
\]
Recall that an operator with deficiency $m$ maps $\cP_n$ to
$\cP_{n-m}$, but not $\cP_{n-m-1}$.  Hence all operators with $m\geq
0$ map $\cP_n$ to $\cP_n$ and can be used to construct an evolution
equation solvable by non-linear separation of variables.
\begin{table}[h]
  \caption{ Quadratic autonomous operators acting on $\mathcal
    P_4$}
  \begin{center}
    \begin{tabular}{c|rrr@{\hskip1em}|ccc}
      $k$ &  &$Q_{k,r}$ && $m$\\
      \hline
      -4 &\vbox to 12pt{}$u_0^2$ &&& -4\\
      -3 &$u_0 u_1$ &&& -3\\
      -2 &$u_1^2$ &  $3u_1^2-4u_0u_2$ && -2 & 0 \\
      -1 &$u_1u_2$ & $u_1u_2-2u_0u_3$ && -1 & 1\\
      0 &$u_2^2$  & $2u_2^2 - 3 u_1u_3$ &$u_2^2-2u_1u_3 + 2 u_0u_4$ &0 &
      2 & 4\\
      1 &$u_2u_3$ &$u_2u_3-3u_1u_4$ && 1 & 3\\
      2 &$u_3^2$ & $u_3^2-2u_2u_4$ && 2& 4\\
      3& $u_3u_4$ &&&3\\
      4&$u_4^2$ &&&4
    \end{tabular}
  \end{center}
\end{table}


\begin{table}[h]
  \caption{ Quadratic autonomous operators acting on $\mathcal
    P_5$}
  \begin{center}
    \begin{tabular}{c|rrr|ccc}
      $k$ &  &$Q_{k,r}$ && $m$\\
      \hline
      -5 & \vbox to12pt{}$u_0^2$ &&& -5 \\
      -4 & $u_0\,u_1$ &&& -4 \\
      -3 & $u_1^2$ & $ 4u_1^2-5u_0u_2$ && -3 & -1\\
      -2 & $u_1u_2$ & $3u_1u_2-5u_0u_3$ && -2 & 0\\
      -1 & $u_2^2$  & $3u_2^2 - 4 u_1u_3$ & $9u_2^2-16u_1u_3 + 10
      u_0u_4$ & -1 & 1 & 3\\
      0 & $u_2u_3$ & $u_2u_3-2u_1u_4$ & $u_2u_3-3u_1u_4+5u_0u_5$ & 0 &
      2 & 4\\
      1 & $u_3^2$ & $2u_3^2-3u_2u_4$ & $u_3^2-2u_2u_4+2u_1u_5$ & 1 & 3
      & 5\\
      2 & $u_3u_4$ & $u_3u_4-3u_2u_5$ && 2 & 4\\
      3 & $u_4^2$ & $u_4^2-2u_3u_5$ && 3 & 5\\
      4 & $u_4u_5$ &&& 4 \\
      5 & $u_5^2$ &&& 5
\end{tabular}
\end{center}
\end{table}


\begin{table}[h]
\caption{ Quadratic autonomous operators acting on $\mathcal
P_6$}
  \small
  \noindent
  \hskip-3em
  \begin{tabular}{c|rrrr|cccc}
    $k$ &  &$Q_{k,r}$ &&& $m$\\
    \hline
    -6 & \vbox to12pt{}$u_0^2$ &&&&-6\\
    -5 & $u_0\,u_1$ &&&&-5\\
    -4 & $u_1^2$ & $5u_1^2-6u_0u_2$ &&&-4&-2\\
    -3 & $u_1u_2$&  $2u_1u_2-3u_0u_3$ &&&-3&-1\\
    -2 & $u_2^2$ & $4u_2^2 - 5 u_1u_3$ & $6u_2^2-10u_1u_3 + 5 u_0u_4$&&-2&0&2\\
    -1 & $u_2u_3$ & $3u_2u_3-5u_1u_4$ & $2u_2u_3-5u_1u_4+5u_0u_5$ && -1&1&3\\
    0 &  $u_3^2$ &  $3u_3^2-4u_2u_4$ &  $9u_3^2-16u_2u_4+10u_1u_5$ &
    $u_3^2-2u_2u_4 + 2u_1u_5-2u_0u_6$ & 0&2&4&6\\
    1 &  $u_3u_4$ &  $u_3u_4-2u_2u_5$ &  $u_3u_4-3u_2u_5+5u_1u_6$ &&1&3&5\\
    2 &  $u_4^2$ &  $2u_4^2-3u_3u_5$ &  $u_4^2-2u_3u_5+2u_2u_6$ &&2&4&6\\
    3 &  $u_4u_5$ &  $u_4u_5-3u_3u_6$ &&&3&5\\
    4 &  $u_5^2$ &  $u_5^2 - 2 u_4u_6$ &&&4&6\\
    5 &  $u_5u_6$ &&&&5\\
    6 &  $u_6^2$ &&&&6
\end{tabular}
\end{table}



\vskip 0.5cm


\end{document}